\documentclass[aps,12pt]{revtex4}
\usepackage{epsfig,amsmath,amssymb}

\newcommand{\eq}{{\, \equiv\, }}
\newcommand{\fr}[1]{
             \frac{#1}}
\newcommand{\bea}{\begin{eqnarray}}
\newcommand{\eea}{\end{eqnarray}}

\newcommand{\chibar}{\overline{\chi}}

\newcommand{\ket}{{\cal i}}
\newcommand{\bra}{{\cal h}}
\newcommand{\gc}{\bra\fr{\alpha_s}{\pi}G^2\ket}

\newcommand{\qc}{\bra\,\overline{q}q\,\ket}

\newcommand{\ga}{{g_{{\cal A}}}}
\newcommand{\gat}{\tilde{g}_{{\cal A}}}

\newcommand{\e}{\varepsilon}
\begin{document}
\title{The $\beta$-term  for $D^* \rightarrow D \, \gamma$ within\\
 a heavy-light chiral quark model}
\author{A. Hiorth and J. O. Eeg}
\email{aksel.hiorth@fys.uio.no}
\email{j.o.eeg@fys.uio.no}
\affiliation{Department of Physics, University of Oslo,
P.O.Box 1048 Blindern, N-0316 Oslo, Norway}


\begin{abstract}
We present a calculation of the $\beta$-term for $D^* \rightarrow D \gamma$
within  a heavy-light chiral quark model. Within the model, soft gluon effects
in terms of the gluon condensate with lowest dimension are included. Also, 
calculations of $1/m_c$ corrections are performed.
We find that the value of $\beta$ is rather sensitive to the
constituent
quark mass compared to other quantities calculated within the same model.
Also, to obtain a value close to the experimental value, one has
to choose a constituent light quark mass larger than for other
quantities studied in previous papers. For a light quark mass in the
range 250 to 300 MeV and a quark condensate in the range 
-(250~-~270~MeV)$^3$ we find the value (2.5 $\pm$ 0.6) GeV$^{-1}$.
This value is in agreement with the value of $\beta$ extracted from
experiment (2.7 $\pm$ 0.2) GeV$^{-1}$.

\end{abstract}

\maketitle

\section{Introduction}
Strong and electromagnetic interactions involving heavy and light mesons
and photons may be described by heavy-light chiral Lagrangians \cite{itchpt}.
Such Lagrangians are determined by chiral symmetry in the light sector in 
addition to heavy quark symmetry. However, the coupling constants of the 
 terms in the Lagrangian are in general unknown.
 In this paper we focus on  the so called $\beta$-term for
 $D^* \rightarrow D \gamma$, which is  studied in \cite{Amund,Stewart,Cheng}.
 This term in the Lagrangian is also relevant for
 calculation of processes like $D^0 \rightarrow 2 \gamma$
\cite{FSZ1}, and $D^0 \rightarrow e^+ e^- \gamma$ \cite{FSZ2}.

Within chiral perturbation theory alone, the value of $\beta$ is not 
determined.
 In this paper we will calculate the $\beta$-term to order $1/m_c$
 within a Heavy-Light Chiral Quark Model (HL$\chi$QM) developed 
recently \cite{ahjoe}.
This model belongs to a class of
   quark loop models \cite{chiqm,Bijnens,barhi} where
the quarks couple directly to the mesons at the scale of chiral
symmetry breaking 
$\Lambda_\chi$ of order  1 GeV. In contrast to most other versions 
of such models, our version 
also incorporates soft gluon effects in terms of the gluon condensate
with lowest dimension
\cite{ahjoe,ahjoeB,EFZ,pider,epb,BEF,EHP}.
At quark level the Lagrangian of our model \cite{ahjoe}
includes the Lagrangian for  
Heavy Quark Effective Field Theory (HQEFT) \cite{neu}. This
 means that $1/m_c$ corrections can be calculated within the model.
It has been problematic to pin down a precise value for $\beta$, and
chiral  loop contributions to $D^* \rightarrow D \gamma$ 
are sizeable \cite{Amund,Stewart,Cheng}.
This may indicate that the chiral expansion will be problematic for $\beta$.

\section{The value of $\beta$ extracted from experiment}

The width of the ${D^{*}}^+$ meson has been measured very accurately\cite{ana} :
\begin{equation}\label{Dwidth}
\Gamma({D^{*}}^+)=(96\pm 26)\text{ keV}\,
\end{equation}
from this value of the width it is possible to extract a value for the
$D^*D\pi$ coupling\cite{ana} 
\begin{equation}\label{gaexp}
\ga^{D^*D\pi}=0.59\pm0.08\; .
\end{equation}
From (\ref{Dwidth}) we find the following radiative decay rate :
\begin{equation}\label{betaexp1}
\Gamma({D^*}^+\to D^+\gamma)=(1.54\pm0.56)\text{ keV}\; ,
\end{equation}
where we have used the branching ratio $Br({D^*}^+\to
D^+\gamma)=(1.6\pm 0.4)\%$\cite{pdg}. To one-loop in the chiral
expansion and to order $1/m_c$ we have \cite{Amund,Cheng}~:
\begin{equation}\label{betaexp2}
\Gamma({D^*}^+\to
D^+\gamma)=\fr{\omega^3\alpha}{3 M_{D^*}}\sqrt{M_{D^*}M_D}
\left[-\fr{1}{3}\beta_d+\fr{2}{3}\fr{1}{m_c}+\fr{\ga^2}{8\pi}\fr{m_\pi}{f^2}\right]^2
\; ,
\end{equation} 
where $f$ is the bare pion decay constant $f=86$ MeV\cite{Cheng}, and
$\omega$ is the photon energy $\omega=(M_{D^*}^2-M_D^2)/(2M_{D^*})$.
The term $1/m_c$ is due to photon emission from the c-quark.
The
bare coupling $\ga$ is related to $\ga^{D^*D\pi}$. In the Heavy-Light
Chiral Quark Model \cite{ahjoe}, described to a certain extent
in the next section, it is estimated to
be $\ga=0.57\pm0.05$. This estimate includes chiral corrections to one
loop and $1/m_Q$ corrections \cite{ahjoe}. This value is very close to
(\ref{gaexp}), in the following we will use the experimental value as
the bare coupling. 
The quantity $\beta_d$ is the effective, renormalized $\beta$
including chiral corrections for
the case $D^+$ when the $D^*$ and $D$ mesons contain
 a $d$-quark \cite{Cheng}:
\begin{eqnarray}
\beta_d=\beta\left\{1+\ga^2\left(\fr{9}{2}\e_\pi+3\e_K+\fr{1}{2}\e_{\eta_8}\right)
+3\left(\e_\pi+\fr{\ga^2}{3}\left(-\fr{3}{2}\e_\pi + 
\e_K+\fr{1}{6}\e_{\eta_8}\right)\right)\right\} \; \, ,
\label{betad}
\end{eqnarray}
where we have used the notation :
\begin{eqnarray}
\e_X=\fr{m_X^2}{32\pi^2 f^2} \ln\fr{\Lambda_\chi^2}{m_X^2} \quad , \, \; \; 
X= \pi, K, \eta_8 \; .
\end{eqnarray}
Eq. (\ref{betad}) includes  vertex corrections for the 
$\beta$-term and field renormalization.
Combining equation (\ref{betaexp1}) and
(\ref{betaexp2}), we find two possible solutions for $\beta_d$ :
\begin{eqnarray}
\beta_d=(0.7\pm0.5)\text{ GeV$^{-1}$} \qquad , \quad
\beta_d=(4.1\pm0.2)\text{ GeV$^{-1}$}\label{b_exp2}\; ,
\end{eqnarray}
where we have used $m_Q=m_c=1.4$ GeV.
These values correspond to the bare values
\begin{eqnarray}
\beta=(0.7\pm0.4)\text{ GeV$^{-1}$}    \qquad , \quad
\beta=(2.7\pm0.2)\text{ GeV$^{-1}$}\label{b_expX}\; .
\end{eqnarray}
 Dropping the chiral corrections
in (\ref{betaexp2}) ($\beta_d=\beta$) would give the result:
\begin{eqnarray}
\beta=-(0.1\pm0.3)\text{ GeV$^{-1}$} \qquad , \quad
\beta=(3.0\pm0.3)\text{ GeV$^{-1}$}\; .
\label{b_exp3}
\end{eqnarray}
The values in (\ref{b_expX}) and (\ref{b_exp3})
 are consistent with those obtained
in \cite{FSZ1}. Both the first values of $\beta$ in equations
(\ref{b_expX}) and (\ref{b_exp3})
 are too low in order to agree with ${D^*}^0$ data \cite{FSZ1}.
Thus, we are left with the second values of
 (\ref{b_expX}) and (\ref{b_exp3}).

\section{The heavy light chiral quark model}

In this section we will give a short description of the Heavy-Light
 Chiral Quark Model (HL$\chi$QM) to be used in this paper.
 The model is based on Heavy Quark Effective Theory
 (HQEFT), which is a systematic expansion in
$1/m_Q$ \cite{neu} (where $m_Q$ is $m_c$ in our case).
 The heavy quark field $Q(x)$ is replaced with a ``reduced''
field, $Q_v(x)$, which is related to the full
 field the in following way:
\begin{equation}
Q_v(x)=P_{+}e^{- im_Q v \cdot x}Q(x) \, ,
\end{equation}
where $P_\pm$ are projection operators $P_\pm=(1 \pm \gamma \cdot v)/2$. The
reduced field $Q_v$  annihilates heavy quarks. 
 The Lagrangian for heavy
quarks is :
\begin{equation}
{\cal L}_{\text{HQEFT}} \, = \, \overline{Q_{v}} \, i v \cdot D \, Q_{v} 
 + \frac{1}{2 m_Q}\overline{Q_{v}} \, 
\left( - C_M \frac{g_s}{2}\sigma \cdot G
 \, +   \, (i D_\perp)_{\text{eff}}^2  \right) \, Q_{v}
 + {\cal O}(m_Q^{- 2})
\label{LHQEFT}
\end{equation}
where $D_\mu$ is the covariant derivative containing the gluon field
(eventually also the photon field), and 
$\sigma \cdot G = \sigma^{\mu \nu} G^a_{\mu \nu} t^a$, where 
$\sigma^{\mu \nu}= i [\gamma^\mu, \gamma^\nu]/2$, $G^a_{\mu \nu}$
is the gluonic field tensors, and $t^a$ are the colour matrices. 
This chromo-magnetic term has a factor $C_M$ which is  one at tree level,
 but slightly modified by perturbative QCD effects below $m_Q$. 
It has been calculated to next to leading order (NLO) \cite{neub,GriFa}.
 Furthermore,  
$(i D_\perp)_{\text{eff}}^2 =
C_D (i D)^2 - C_K (i v \cdot D)^2 $. At tree level, $C_D = C_K = 1$.
Here, $C_D$ is not modified by perturbative QCD, while $C_K$ is different 
from one due to perturbative QCD corrections \cite{GriFa}.

The Lagrangian for the HL$\chi$QM is 
\begin{equation}
{\cal L}_{\text{HL$\chi$QM}} =  {\cal L}_{\text{HQEFT}} + 
 {\cal L}_{\chi\text{QM}}  +   {\cal L}_{\text{Int}} \; .
\label{totlag}
\end{equation}
The first term is given in equation (\ref{LHQEFT}) (Note however, that
only soft gluons are considered to be included in (\ref{totlag}) )~.
The light quark sector is described by the Chiral Quark Model 
($\chi$QM) \cite{chiqm},
having a standard QCD term and a term describing interactions between
quarks and  (Goldstone) mesons: 
\begin{equation}
{\cal L}_{\chi\text{QM}} =  
\chibar \left[\gamma^\mu (i D_\mu   +    {\cal V}_{\mu}  +  
\gamma_5  {\cal A}_{\mu})    -    m \right]\chi 
  -     \chibar \widetilde{M_q} \chi \;  .
\label{chqmR}
\end{equation}
Here $m$ is the (SU(3)-invariant) constituent light quark mass and 
$\chi$ is the flavour rotated quark fields given by:
$\chi_L  =   \xi^\dagger q_L$ and $\chi_R  =   \xi q_R$,   
where $q^T  =  (u,d,s)$ are the light quark fields. The left- and
 right-handed
 projections $q_L$ and $q_R$ transforms after $SU(3)_L$ and $SU(3)_R$
respectively.
The quantity $\xi$ is a 3 by 3 matrix containing
 the (would be)  Goldstone octet ($\pi, K, \eta$).
In terms of $\xi$
the vector and axial vector fields
${\cal V}_{\mu}$ and  
${\cal A}_\mu$ in (\ref{chqmR})  are given by:
\begin{equation}
{\cal V}_{\mu}\eq \fr{i}{2}(\xi^\dagger\partial_\mu\xi
+\xi\partial_\mu\xi^\dagger 
) \quad ,  \; \; \;
{\cal A}_\mu\eq  -  \fr{i}{2}
(\xi^\dagger\partial_\mu\xi
-\xi\partial_\mu\xi^\dagger) \quad , \; \; \;  \xi \equiv \exp{(i\Pi/f)}\,
\label{defVA}
\end{equation}
where $f$ is the bare pion coupling, and $\Pi$ is a  3 by 3 matrix
which contains the Goldstone bosons $\pi,K,\eta$ in the standard way.
In (\ref{chqmR}) the quantity $\widetilde{M}_q$
contains the
 current quark mass matrix ${\cal M}_q$ and and the Goldstone fields
through $\xi$:
\bea
&&\widetilde{M_q} \eq \widetilde{M}_q^V   +    \widetilde{M}_q^A \gamma_5  \;
 , \; \text{where} \label{cmass}\\
\widetilde{M}_q^V \, \eq \, 
\fr{1}{2}(\xi^\dagger {\cal M}_q^\dagger\xi^\dagger \,
+&& \xi {\cal M}_q\xi )\quad\text{and}\quad 
\widetilde{M}_q^A\eq -\fr{1}{2}(\xi^\dagger {\cal M}_q^\dagger\xi^\dagger
 -\xi {\cal M}_q\xi) \; .
\label{masst}
\eea

The interaction between heavy meson fields and heavy quarks are
described by the following Lagrangian :
\begin{equation}
{\cal L}_{Int}  =   
 -   G_H \, \left[ \chibar_a \, \overline{H_a} 
\, Q_{v} \,
  +     \overline{Q_{v}} \, H_a \, \chi_a \right] \; ,
\label{Int}
\end{equation}
where $G_H$ is a  coupling constant, and
 $H_a$ is the heavy meson field  containing
 a spin zero and spin one boson ($a$ is a SU(3) flavour index):
\begin{eqnarray}
&H_a & \eq  P_{+} (P_\mu^a \gamma_\mu -     
i P_5^a \gamma_5)\; \; \; , \; \; 
\overline{H_a}
\eq  \gamma^0 (H_a)^\dagger \gamma^0 \; .
\label{barH}
\end{eqnarray}
The fields $P^a$ annihilates  a heavy meson containing
 a heavy quark with velocity  $v$. 

Integrating out the quarks by using by using (\ref{LHQEFT}), (\ref{chqmR})
and  (\ref{Int}), the lowest order chiral Lagrangian up
to ${\cal O}(m_Q^{-1})$ can be written as 
\cite{itchpt,ahjoe}
\begin{equation}
{\cal L} \, = \, - \, Tr\left[\overline{H_a}
iv\cdot {\cal D}_{ba}
H_b\right]\, -\, 
\ga \, Tr\left[\overline{H_a} H_b
\gamma_\mu\gamma_5 {\cal A}^\mu_{ba}\right]\, 
,\label{LS1}
\end{equation}
where $i{\cal D}^\mu_{ba}=i \delta_{ba} D^\mu-{\cal V}^\mu_{ba}$
and the axial coupling $\ga$ is of order 0.6.
 Eqs. (\ref{LS1}) and (\ref{defVA}) will be
used for the chiral loop contributions.
(Note that the eqs. (\ref{Int}) and (\ref{LS1}) both contain an
 additional term \cite{ahjoe} .
These are, however,  irrelevant in the present paper).

To obtain (\ref{LS1}) from the HL$\chi$QM  one encounters divergent loop
integrals, which will  be  quadratic-, linear- and
logarithmic divergent. Thus $G_H^2$ times linear combinations of these 
divergent integrals have to be put equal to 1 (the normalization) and
$\ga$ respectively. For details, see the Appendix. 
Within our model,
the values for the regularized  versions of the quadratic-, linear-,
and
logarithmic divergent integrals  are determined
 by the physical values of $\qc$,  $\ga$, and  $f$ respectively.
 The effective coupling $G_H$ describing the
interaction between the quarks and heavy mesons can be expressed in terms
of $m$, $f$, $g_{\cal A}$, and the  mass splitting between
 the $1^-$ state and $0^-$ state.
Using the equations (\ref{norm}), (\ref{ga}), (\ref{I2}), and 
 (\ref{gh}) in the Appendix,
 one finds the following  relations between this mass-splitting and the
gluon condensate   via the chromomagnetic 
interaction in (\ref{LHQEFT}) \cite{ahjoe} :
\begin{equation}
\label{gcgh}
\gc  =   \fr{16 f^2}{\pi \eta} \, \fr{\mu_G^2}{\rho}  \; \, , \qquad  
G_H^2  =   \fr{2m}{f^2} \, \rho \; \, ,
\qquad  \eta \, \eq\, 
\fr{(\pi +  2)}{\pi}C_M(\Lambda_\chi) \; \, ,
\end{equation}
where 
\begin{equation}
\label{rho}
\rho \, \eq \, \fr{(1 +  3g_{\cal A}) + 
 \frac{\mu_G^2}{\eta \, m^2}}{4 (1 + \frac{N_c m^2}{8 \pi f^2} )}
\; \; \; , \; \; \; \mu_G^2(H) \, =  \, \frac{3}{2} m_Q (M_{H^*} -  M_H) .  
\end{equation}

When calculating the soft gluon effects in terms of the gluon condensate,
we follow  \cite{nov}.
The gluon condensate is obtained by the replacement :
\begin{equation}
g_s^2 G_{\mu \nu}^a G_{\alpha \beta}^b  \; \rightarrow \fr{4 \pi^2}{(N_c^2-1)}
\delta^{a b} \gc \frac{1}{12} (g_{\mu \alpha} g_{\nu \beta} -  
g_{\mu \beta} g_{\nu \alpha} ) \, .
\label{gluecond}
\end{equation}
Soft gluons
coupling to a heavy quark is suppressed by $1/m_Q$, since to leading
order the vertex is proportional to $v_\mu v_\nu G^{a\mu\nu}= 0$,
 $v_\mu$ being the heavy quark velocity. 
We observe that the mass-splitting between $H$ and $H^*$
sets the scale of the gluon condensate.
Within the pure light sector, 
the  gluon condensate contribution obtained by using (\ref{gluecond})
has apriori an arbitrary value. 
One  may assume that it has the same value as in QCD sum rules,
or it has to  be fitted to some
effect, in our case the mass splitting between pseudoscalar and vector heavy
mesons.

The $1/m_Q$ corrections to the strong Lagrangian have been calculated in
\cite{ahjoe}. They may  formally be put 
into spin dependent renormalization factors.
This means  that (\ref{LS1}) is still valid with the replacement
$H \rightarrow  H^{\text{r}}= H \, (Z_H)^{-\frac{1}{2}}$, where $Z_H$
 and the renormalized (effective) coupling $\gat$ are defined as:
\bea
Z_H^{-1}  &=& 1+\fr{\varepsilon_1-2d_M\varepsilon_2}{m_Q} \; \;  , \\
\gat &=& \ga \left(1-\fr{1}{m_Q}(\varepsilon_1-2d_{\cal A}\varepsilon_2)\right)
-\fr{1}{m_Q}(g_1-d_{\cal A}g_2) \; \, ,
\eea
where
\begin{equation}
d_M=\begin{cases}\,\,\,\,\, 3\quad\text{for}\quad 0^-\\
                 -1\quad\text{for}\quad 1^-
\end{cases}\quad\quad
d_{\cal A}=\begin{cases}\,\,\,\, 1\quad\text{for}\quad H^*H\,\,\,\quad\text{coupling}\\
              -1\quad\text{for}\quad H^*H^*\quad\text{coupling}
\end{cases}
\end{equation}
and :
\bea
\varepsilon_1&=&-m+G_H^2\left(\fr{\qc}{4m}+f^2+\fr{N_cm^2}{16\pi}+\nonumber
\fr{C_K}{16}(\fr{\qc}{m}-f^2)\right.\\&&\left.\qquad\qquad\quad
+\fr{1}{128 m^2}(C_K+8-3\pi) \gc\right) \; , \\
g_1&=&\,m-G_H^2\left(\fr{\qc}{12m}+\fr{f^2}{6}+
\fr{N_cm^2(3\pi+4)}{48\pi}- \nonumber
\fr{C_K}{16}(\fr{\qc}{m}+3f^2)\right.\\&&\left.\qquad\qquad\quad
+\fr{1}{64 m^2}(C_K-2\pi)\gc\right) \; , \\
g_2&=& \fr{(\pi + 4)}{(\pi+2)} \fr{\mu_G^2}{6 m}\;,\qquad \e_2=-\fr{g_2}{2}
\; .
\eea

\section{The $\beta$ term for $D^*\to D\gamma$}

\begin{figure}
\epsfig{file=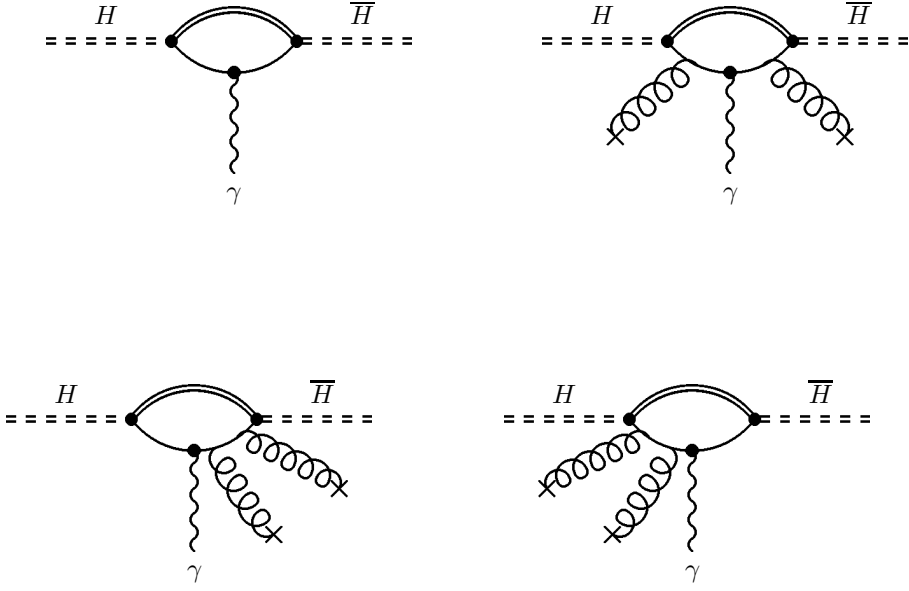}
\caption{Leading order diagrams contributing to $\beta$.
The double and single lines represent heavy and light quarks,
respectively.
The double dashed lines represent heavy mesons, and the wavy
lines represent emission of soft gluons ending in vacuum to make 
gluon condensates.}
\label{fig:beta_LO}
\end{figure}

The chiral Lagrangian $\beta$-term has the form \cite{Stewart}:
\begin{equation}
{\cal L}_{\beta}=\fr{e \beta}{4}Tr[\overline{H} \, H \, \sigma\cdot
F \, Q^\xi]\, ,
\label{beta}
\end{equation}
where $Q^\xi=(\xi^\dagger Q\xi+\xi Q\xi^\dagger)/2$,
$Q=diag(-2/3,-1/3,-1/3)$, and $F$ is the electromagnetic filed tensor.
The $\beta$ term can be calculated in 
HL$\chi$QM, by considering the diagrams in figure~\ref{fig:beta_LO}.
These diagrams will be calculated following  
\cite{nov}, both for the gluon and the electromagnetic field.
The first diagram without gluons is logarithmically divergent.
Performing the calculation of the rest of the diagrams we used the algebraic 
program  FORM \cite{FORM}. We 
obtained the following expression :
\begin{equation}
\label{betaLOB}
\beta_{LO} =
\fr{G_H^2}{2}\left\{-4 i N_c I_2 +\fr{N_c }{4\pi}
-\fr{1}{4 m^4}\left(\fr{32+3\pi}{144}\right)\gc\right\}\, .
\end{equation}
Using the relation (\ref{I2}) for the logarithmic divergent integral 
$I_2$, we obtain:
\begin{equation}\label{betaLO}
\beta_{LO} =
\fr{G_H^2 f^2}{2m^2}\left\{1+\fr{N_c m^2}{4\pi f^2}
-\fr{1}{4f^2 m^2}\left(\fr{56+3\pi}{144}\right)\gc\right\}\, .
\end{equation}

\begin{figure}
\epsfig{file=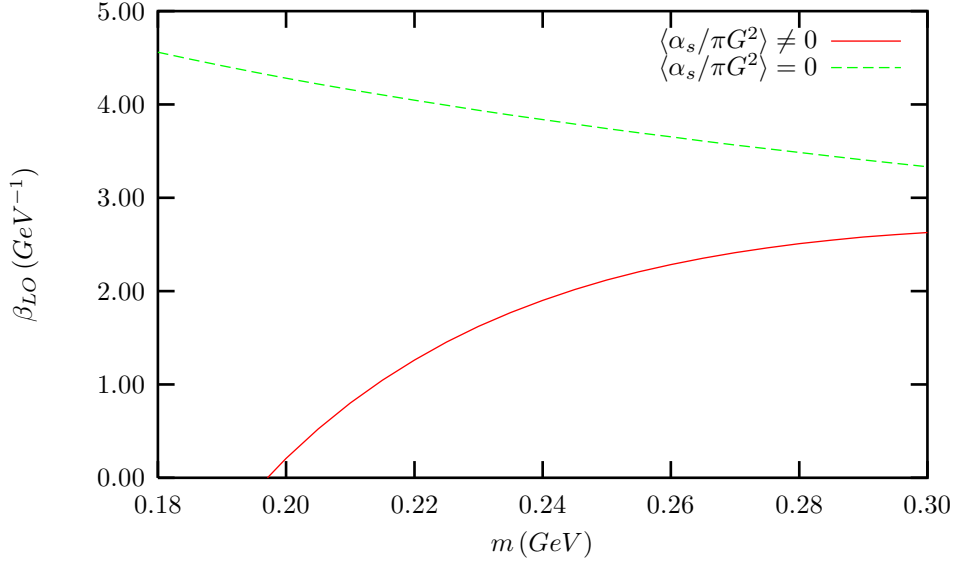}
\caption{$\beta_{LO}$ as a function of the constituent light quark mass
for parameters from ref. \cite{ahjoe}}
\label{fig:betap}
\end{figure}

As seen from figure~\ref{fig:betap}, $\beta$ depends strongly on the
constituent light quark mass $m$, especially for $m$ below 250 MeV.
 There is a  
partial   cancellation between large terms in
(\ref{betaLO}),
 and for
 values of $m$ smaller than about 200 MeV,
 $\beta_{LO}$ turns negative. Note also that if the
gluon condensate is dropped in (\ref{betaLO}), $\beta$ would be too
big, as seen from figure~\ref{fig:betap}.

\begin{figure}
\epsfig{file=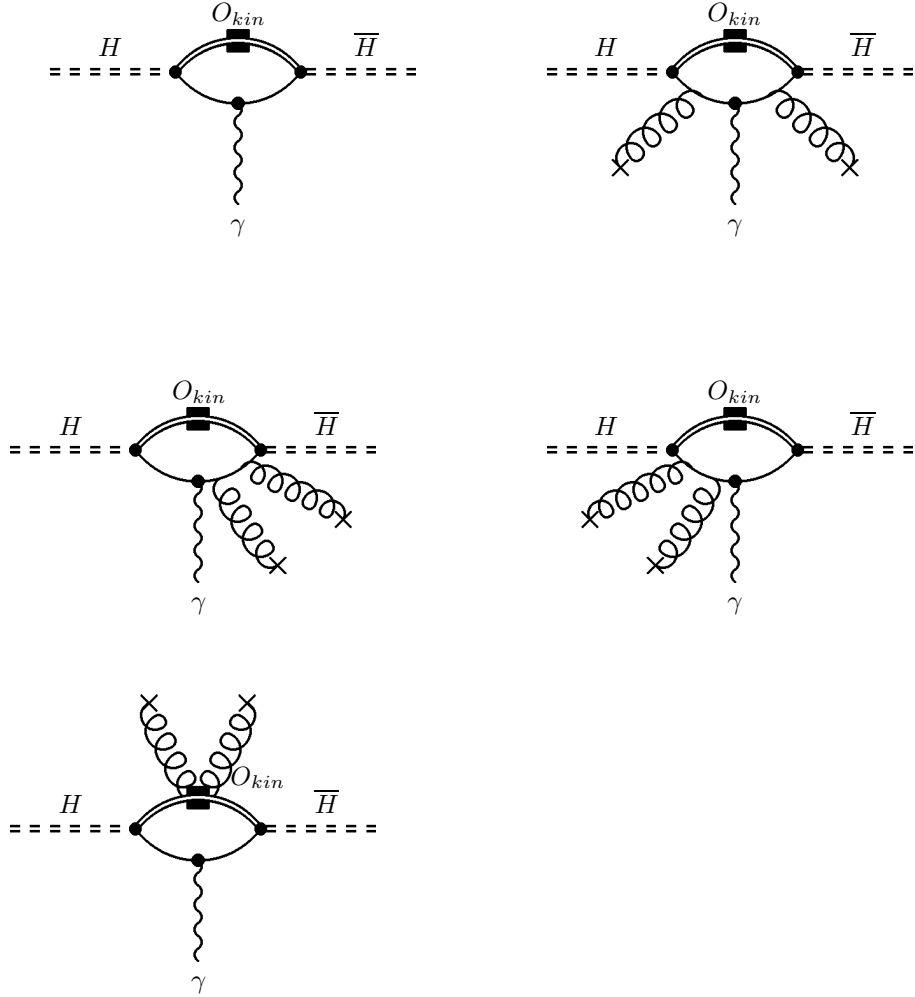}
\caption{$1/m_Q$ diagrams from the kinetic operator, contributing to $\beta$}
\label{fig:beta_MQ1}
\end{figure}

\begin{figure}
\epsfig{file=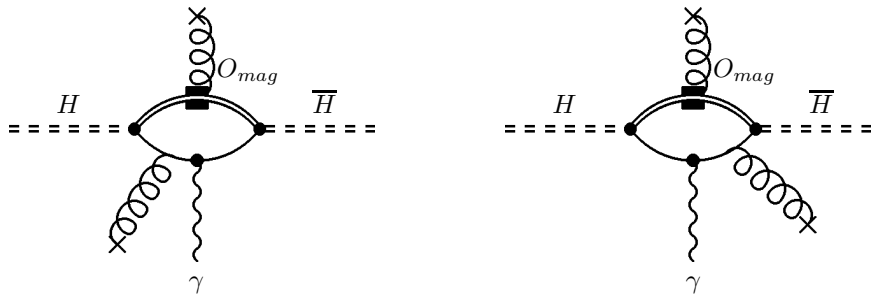}
\caption{$1/m_Q$ diagrams from the chromo magnetic  operator, 
contributing to $\beta$}
\label{fig:beta_MQ2}
\end{figure}

Apriori one might hope that
$1/m_Q$ corrections stabilizes $\beta$ for values of $m$ in the range
190 to 250 MeV used in \cite{ahjoe}. However, as seen 
in the following, this will not be the case. 
The $1/m_Q$ corrections to $\beta$ can be found by calculating the
diagrams in figures~\ref{fig:beta_MQ1} and \ref{fig:beta_MQ2}.  
We find that the kinetic term in (\ref{LHQEFT}) 
give a negative contribution to  $\beta$,
and the chromomagnetic term a positive contribution. However, the kinetic 
term dominates in absolute value, and in total, 
$1/m_Q$ corrections give a negative contribution to $\beta$ for $m$
below 250 MeV.

\begin{figure}
\epsfig{file=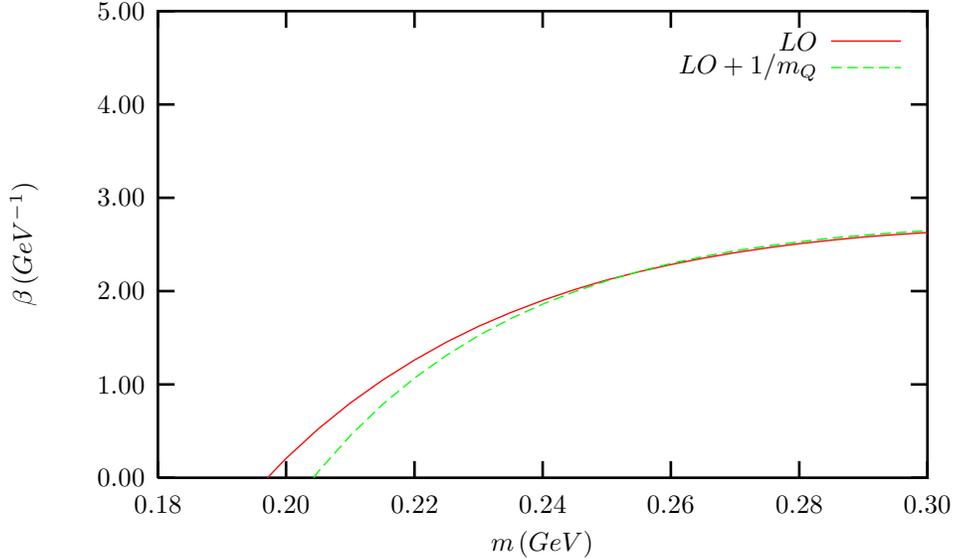}
\caption{$\beta$ as a function of the constituent  light quark mass for 
$\ga = 0.59$}
\label{fig:betaga}
\end{figure}

Combining the leading order and  the $1/m_Q$ corrections  we find :
\bea
&\beta=&\sqrt{Z_H Z_{H^*}}\beta_{LO}+\fr{\tau_\beta}{m_Q}   \; \, ,
\; \, \text{where}    \nonumber \\
&\tau_\beta=&-\fr{3}{4}(1-\ga) - \fr{G_H^2}{4 m} \left\{
2 f^2-\fr{m^2 N_c}{2\pi^2} +
\left[\fr{1}{12}+\fr{\pi}{192}\nonumber\right.\right.\\ &&\left.\left.-
C_M\fr{(\pi+4)}{72}\right]\fr{1}{m^2}\gc - 
C_K\left(f^2
-\fr{1}{36 m^2}\gc\right)\right\} \; \, .
\eea
where $\tau_\beta$ contains the result of the diagrams in the 
figures~\ref{fig:beta_MQ1} and \ref{fig:beta_MQ2}.
A plot of $\beta$ is shown in figure~\ref{fig:betaga}. As we see, the result 
varies significantly with $m$. This is in contrast to other quantities
studied in \cite{ahjoe,ahjoeB}. We observe that  the quantities
studied in \cite{ahjoe,ahjoeB} have contributions of zeroth order
in $m$. In contrast, 
$\beta$ is of order $1/m$, which means that $\beta$
 is expected to be more sensitive to variations of $m$.
The $1/m_Q$ corrections are not sizeable, even if the charm quark 
 is not very heavy. However, they pull in the wrong direction compared to
 the experimental value of order 2-3 GeV$^{-1}$.
To obtain a value closer to the experimental value for $\beta$,
 we need a  value
for $m$ higher than used in \cite{ahjoe,ahjoeB}. However, this will lead to 
values of $f_B$ and
$f_D$ which are too small .
 This can be compensated by  using a higher value of the quark
condensate
$\qc$ than the one obtained in QCD sum rules, which was the one 
 used in \cite{ahjoe,ahjoeB}. This might be acceptable because it is not clear
that our model dependent quantities $\qc$ and  $\gc$ should be
 exactly those obtained
in QCD sum rules. Taking a higher quark condensate and making a new
fit
we obtain the result given in the tables I and II. 
Note that the curve for $\beta$ itself as a function of $m$ does not
change, but we are now allowed to go to higher values for $m$ at the curve.

\begin{table}[t]
\begin{center}
\caption{Predictions of HL$\chi$QM for different input parameters in
the $D$-sector.}

\begin{tabular}{l l| l}\hline\hline
&$\ga^{D^*D\pi}=\;\;\,\,0.59\pm 0.08$
\cite{ana}&$\ga^{D^*D\pi}=\;\;\,\,0.59\pm 0.08$ \cite{ana} \\         
&$\qc\;\,=-(230-250)^3$ MeV&$\qc\;\,=-(250-270)^3$ MeV\\
&$m\quad\;\;\, =\;\;\,(190-250)$ MeV  &$m\quad\;\;\,=\;\;\,(250-300)$ MeV  \\ \hline
$G_D$           & $\quad(7.2\pm 0.6)\,$ GeV$^{-1/2}$& $\quad(6.4\pm 0.4)\,$ GeV$^{-1/2}$ \\
$\gc^{1/4}$     & $\quad(290\pm 20)\,$ MeV     & $\quad(330\pm 20)\,$ MeV \\
$g_1$           & $\quad(0.8\pm 0.1)\,$ GeV           & $\quad(0.61\pm 0.07)\,$ GeV \\
$g_2$           & $\quad(0.32\pm 0.04)\,$ GeV         & $\quad(0.25\pm 0.02)\,$ GeV \\
$\e_1$          & $-(0.8\pm 0.3)\,$ GeV          & $-(0.6\pm 0.2)\,$ GeV \\
$\lambda_1$     & $\quad 0.8\pm 0.2$                 & $\quad 0.5\pm 0.1$\\
$\mu_\pi^2$     & $\quad(0.34\pm 0.05)\,$ GeV$^2$   & $\quad(0.29\pm 0.04)\,$ GeV$^2$ \\
$f_D$           & $\quad(220\pm 60)\,$ MeV          & $\quad(220\pm 35)\,$ MeV\\
$f_{D^*}$       & $\quad(260\pm 85)\,$ MeV          & $\quad(235\pm 50)\,$ MeV \\
$f_{D_s}$       & $\quad(245\pm 70)\,$ MeV          & $\quad(260\pm 45)\,$ MeV \\
$f_{D^*_s}$     & $\quad(280\pm 95)\,$ MeV          & $\quad(270\pm 60)\,$ MeV \\
$f_{D^*}/f_D$   & $\quad 1.2\pm 0.1$               & $\quad 1.12\pm 0.04$\\
$f_{D_s}/f_D$   & $\quad 1.18\pm 0.06$               & $\quad 1.26\pm 0.04$ \\
$\beta$  & $\quad (0\pm 2)$ GeV$^{-1}$      & $\quad (2.5\pm0.6)$ GeV$^{-1}$\\
\hline\hline
\end{tabular}
\end{center}
\end{table}

\begin{table}[t]
\begin{center}
\caption{Predictions of HL$\chi$QM for different input parameters in
the $B$-sector.}
\begin{tabular}{l l| l}\hline\hline
&$\ga^{D^*D\pi}=\;\;\,\,0.59\pm 0.08$
\cite{ana}&$\ga^{D^*D\pi}=\;\;\,\,0.59\pm 0.08$ \cite{ana} \\         
&$\qc\;\,=-(230-250)^3$ MeV&$\qc\;\,=-(250-270)^3$ MeV\\
&$m\quad\;\;\, =\;\;\,(190-250)$ MeV  &$m\quad\;\;\,=\;\;\,(250-300)$ MeV  \\ \hline
$G_B$           & $\quad(8.3\pm 0.7)\,$ GeV$^{-1/2}$& $\quad(7.2\pm 0.5)\,$ GeV$^{-1/2}$ \\
$\gc^{1/4}$     & $\quad(300\pm 25)\,$ MeV     & $\quad(340\pm 20)\,$ MeV \\
$g_1$           & $\quad(1.4\pm 0.3)\,$ GeV           & $\quad(1.0\pm 0.2)\,$ GeV \\
$g_2$           & $\quad(0.39\pm 0.05)\,$ GeV         & $\quad(0.31\pm 0.03)\,$ GeV \\
$\e_1$          & $-(0.9\pm 0.4)\,$ GeV          & $-(0.6\pm 0.2)\,$ GeV \\
$\lambda_1$     & $\quad 1.1\pm 0.3$                 & $\quad 0.8\pm 0.1$\\
$\mu_\pi^2$     & $\quad(0.42\pm 0.03)\,$ GeV$^2$   & $\quad(0.39\pm 0.03)\,$ GeV$^2$ \\
$f_B$           & $\quad(190\pm 50)\,$ MeV          & $\quad(185\pm 30)\,$ MeV\\
$f_{B^*}$       & $\quad(200\pm 60)\,$ MeV          & $\quad(190\pm 35)\,$ MeV \\
$f_{B_s}$       & $\quad(210\pm 70)\,$ MeV          & $\quad(215\pm 45)\,$ MeV \\
$f_{B^*_s}$     & $\quad(220\pm 70)\,$ MeV          & $\quad(215\pm 45)\,$ MeV \\
$f_{B^*}/f_B$   & $\quad 1.07\pm 0.04$               & $\quad 1.02\pm 0.02$\\
$f_{B_s}/f_B$   & $\quad 1.14\pm 0.07$               & $\quad 1.22\pm 0.02$ \\
$\hat{B}_{B_d}$ & $\quad 1.51\pm 0.09$                 & $\quad 1.52\pm 0.07$\\ 
$\hat{B}_{B_s}$ & $\quad 1.4\pm 0.1$                   & $\quad 1.4\pm 0.1 $\\
$\xi=\fr{f_{B_s}\sqrt{\hat{B}_{B_s}}}{f_{B_d}\sqrt{\hat{B}_{B_d}}}$
                & $\quad 1.08\pm 0.07$                 & $\quad 1.16\pm 0.04$ \\
$\beta$  & $\quad-(2\pm 3)$ GeV$^{-1}$      & $\quad (1.2\pm0.8)$ GeV$^{-1}$\\
\hline\hline
\end{tabular}
\end{center}
\end{table}

Chiral Lagrangian terms with extra  current quark mass can
 be obtained by
taking the derivative of the expression for $\beta$ in (\ref{betaLOB})
with respect to $m$, when keeping $G_H$ and $\gc$ fixed.
This gives: 
\begin{equation}
{\cal L}_{FM}=\fr{e \widetilde{\alpha}}{8} 
Tr \left[ \overline{H} \, H \, \sigma\cdot
F \, \left(Q^\xi \, \widetilde{M}_q^V + 
\widetilde{M}_q^V Q^\xi \right) \right] \; ,
\end{equation}
where 
\begin{equation}\label{alphat}
 \widetilde{\alpha}=
\fr{G_H^2}{4m}\left\{-\fr{N_c }{\pi^2 }
+ \left(\fr{32+3\pi}{72m^4}\right)\gc\right\}\, .
\end{equation}
This term is  unimportant for $D^* \to D \gamma$ because of the
small current quark masses for  the $u, d$ quarks, but it gives a
sizeable contribution to
 $D_s^* \to D_s \gamma$.

\section{Conclusion}

We have calculated  the quantity $\beta$  and found that it
depends significantly on the 
constituent light quark mass $m$.  
 To obtain a value close to the experimental  one
has to pick values on the high side ($m$= 250 to 300 MeV, say)
compared to the values in \cite{ahjoe,ahjoeB}, where $m= 220 \pm 30$
MeV were used. We  have redone the fits in  \cite{ahjoe,ahjoeB}
at the price of a higher value of the quark condensate $\qc$
in the range $- ($250 to 270 MeV$)^4$.
The results are given in the tables I and II.
For the case $D^* \rightarrow D \gamma$ we have found that
(the bare) $\beta=(2.5 \pm 0.6)$ GeV$^{-1}$ to be compared with
$\beta= (2.7 \pm 0.20)$ GeV$^{-1}$ extracted from experiment.
 
\vspace{0.3cm}
{\bf \flushleft Acknowledgments}\linebreak
We thank Svjetlana Fajfer for fruitful discussions. J.O.E. is
 supported in part by the Norwegian research council and  by
 the European Union RTN
network, Contract No. HPRN-CT-2002-00311  (EURIDICE)

\appendix
\section{The limit $m \rightarrow 0$}
\label{chiral}

In this Appendix we will discuss the limit of restauration of chiral
symmetry, i.e. the limit $m \rightarrow 0$. In order to do this, we
have to consider the various constraints obtained when constructing the
HL$\chi$QM \cite{ahjoe}.


To obtain (\ref{LS1}) from the HL$\chi$QM  one encounters divergent loop
integrals, which will in general be quadratic-, linear- and
logarithmic divergent.
For the kinetic term we obtain the identification:
\begin{equation}
 -  iG_H^2N_c \, \left(I_{3/2}  +   2mI_2
+ i\fr{(3 \pi -8)}{384 m^3 N_c}\gc \right) =   1 \; ,
\label{norm}
\end{equation}
and for  the axial vector term 
\begin{equation}
 -  iG_H^2N_c \, \left( - \frac{1}{3} I_{3/2} + \fr{im}{12\pi}  +   2mI_2    
+ i\fr{(3 \pi -8)}{384 m^3 N_c}\gc \right) =   \ga \; .
\label{gano}
\end{equation}
Combining (\ref{norm}) and (\ref{gano}), the strong axial coupling
 $\ga$ may be written:
\begin{equation}
\ga \, = \, 1 \, - \, \delta \ga \; \; \; , \; \text{where} \; \; \;
 \delta \ga \, = \,
- \, \fr{4}{3}iG_H^2N_c \left(I_{3/2}-\fr{im}{16\pi}\right) \; \, .
\label{ga}
\end{equation}
Here the term 1 for $\ga$ corresponds to the normalization
(\ref{norm}) and the term $\delta \ga$ is a dynamical deviation from
this. It should be noted that also within other models
\cite{CoFaNa,DoXu},
 $\ga$ may be written as $\ga \, = \, 1 \, - \, \delta \ga$,
 but the expression for $\delta \ga$ is model dependent .
We observe that the  (formally) linear divergent
 integral $I_{3/2}$ is related to 
the strong axial coupling $\ga$.  
Analogously,
within the pure light quark sector (the $\chi$QM), it is well known
that the quadratic and
logarithmic divergent integrals are related to the quark condensate and
$f$, respectively \cite{chiqm,pider,epb,BEF}:
\bea
&\qc &= -4imN_cI_1-\fr{1}{12m}\gc \; \, ,\label{I1}\\
&f^2 &=    -  i4m^2N_cI_2 +  \fr{1}{24m^2}\gc \; \, .
\label{I2}
\eea

Eliminating $I_{3/2}$ from the 
eqs. (\ref{norm}) and (\ref{ga}) 
and inserting the expression for $I_2$ obtained from (\ref{I2}) 
we find  the following expression for
$G_H$ :  
\begin{equation}
G_H^2 =  \fr{m(1 +  3g_{\cal A})}{2f^2 +   \fr{m^2 N_c}{4\pi} -  
\frac{\eta_1}{m^2}\gc}\, ,
\quad\text{where}\quad
\eta_1\eq \fr{\pi}{32} \, .
\label{gh}
\end{equation}

The divergent integrals $I_1,\, I_2$ and $I_{3/2}$ are (for $N=1,2$): 
\bea
I_N \, 
\eq 
\,\int\fr{d^dk}{(2\pi)^d}\fr{1}{(k^2 -  m^2)^N}   \; , \, \;
I_{3/2}\,
\eq
\, \int\fr{d^dk}{(2\pi)^d}\fr{1}{(v\cdot k)(k^2 -  m^2)} \; \; ,
\eea
where $I_{3/2}$ is proportional to the cut-off in primitive cut-off 
regularization:
\bea
I_{3/2}= i \frac{\Lambda}{16 \pi} \left(1 + 
{\cal O}(\frac{m}{\Lambda})\right) \; \, , 
\eea
where the cut-off $\Lambda$ (which we only have used in the
qualitative considerations in this Appendix) is of the same order 
as the chiral symmetry breaking scale $\Lambda_\chi$.
In contrast, $I_{3/2}$ is finite and proportional to $m$
in dimensional regularization.
 Within our model,
the values for the regularized  integrals $I_1,\, I_2$ and
 $I_{3/2}$ are determined
 by the numerical values of $\qc$, $f$ and $\ga$, respectively.

Looking at the equations (\ref{I1}) and (\ref{I2}), one may worry \cite{EdR}
that $\qc$ and $f$ behaves like $1/m$ in  the  limit $m \rightarrow
0$ unless one assumes that $\gc$ also go to zero in this limit.
We should stress that the exact limit $m = 0$ cannot be
taken because our loop integrals will then be meaningless. Still we may let
$m$ approach zero without going to this exact limit.
In the pure light sector (at least when vector mesons are not
included)
 there are no restrictions on how $\gc$
might go to zero. In the heavy light sector we have in addition to
(\ref{I1}) and (\ref{I2}) also the relations (\ref{norm}),
(\ref{ga}), and  (\ref{gh})
 which put restrictions on
the behavior of the gluon condensate $\gc$ for small masses.
As $\gc$ has dimension mass to the fourth power, we find that 
$\qc$ and $f^2$ may
go to zero if $\gc$ goes to zero as $m^4$ or $m^3 \Lambda$ (eventually
combined with $\ln (m/\Lambda)$). However, the behavior $m^3 \Lambda$ is
inconsistent with the additional equations 
(\ref{gcgh}) and  (\ref{rho}).
Still, from all equations (\ref{norm})- (\ref{gh}) and 
(\ref{gcgh}), (\ref{rho}), we find the  possible solution
\bea
\gc = \hat{c} \, N_c m^4 K(m) \; \; \text{, where}\quad 
K(m) \equiv (-4i I_2 + \frac{1}{8 \pi}) \; ,
\label{chilim}
\eea
and $\hat{c}$ is some constant.
Then we must have the following behavior for $G_H^2$, $\ga$ and $\mu_G^2$
when $m$ approaches zero:
\bea
G_H^2 \sim \frac{1}{N_c \Lambda} \quad , 
 \; (1+ 3\ga)  \sim \frac{m}{\Lambda} K(m) \;
\quad , \; \; \mu_G^2 \sim  \frac{m^3}{\Lambda} K(m) \; \, ,
\label{chilimR}
\eea
with some restrictions on the proportionality factors.
Here, the regularized $I_2$ is such that for small $m$,  $K(m) = (c_1
+c_2 \ln m/\Lambda)$, $c_1$ and $c_2$ being constants.
 The behavior of  $G_H^2$ is in
 agreement with Nambu-Jona-Lasinio 
models \cite{Bijnens}. Note that in our model, $\delta \ga \rightarrow
4/3$ (corresponding to $\ga \rightarrow
-1/3$) for  $m \rightarrow 0$, in contrast to 
$\delta \ga \rightarrow 2/3$ in \cite{CoFaNa} for a free  Dirac
particle with $m=0$.
Note that in \cite{ahjoe} we gave the variation of the gluon
condensate with $m$ for a fixed value of $\mu_G^2$. For the
considerations in this Appendix,
 we have to let  $\mu_G^2$  go to zero with $m$
in order to be consistent. When $m \rightarrow 0$, we also find that 
$\beta \rightarrow 1/\Lambda$, provided that the coefficient $\hat{c}$
in (\ref{chilim}) is fixed to a specific value ( which is
$\hat{c}= 576/(3 \pi +32) \simeq (1.93)^4$).

\vspace{0.1cm}
Before  closing this section, we will shortly comment that
in the limit where only the leading logarithmic integral $I_2$ is
kept \cite{ahjoe}, when $g_{\cal A} \rightarrow \, 1$ \cite{itchpt}
 and $\rho \rightarrow \, 1$, we obtain from  equation (\ref{betaLO})
 the non relativistic quark model result \cite{Stewart,itchpt} :
\begin{equation}
\beta \rightarrow \beta_{\text{NR}}=\fr{1}{m} \; .
\end{equation}

\newpage

\bibliographystyle{unsrt}

\end{document}